\documentclass{PoS}

\usepackage[utf8]{inputenc} 
\usepackage[T1]{fontenc}

\title{Sensitivity of D meson azimuthal anisotropies to system size and nuclear structure}

\ShortTitle{D meson $v_n\{m\}$ }

\author{Roland Katz\\SUBATECH, Universit\'e de Nantes, EMN, IN2P3/CNRS, 44307 Nantes, France}
\author{Caio A.~G.~Prado\\Institute of Particle Physics, Central China Normal University (CCNU), Wuhan, Hubei 430079, China}

\author{\speaker{Jacquelyn Noronha-Hostler}\\
Department of Physics and Astronomy, Rutgers University, Piscataway, NJ 08854, USA\\
\email{jacquelyn.noronhahostler@phys.rutgers.edu}}
\author{Jorge Noronha\\Instituto de F\'{i}sica, Universidade de S\~{a}o Paulo, C.P. 66318, 05315-970 S\~{a}o Paulo, SP, Brazil}
\author{Alexandre A.~P.~Suaide\\Instituto de F\'{i}sica, Universidade de S\~{a}o Paulo, C.P. 66318, 05315-970 S\~{a}o Paulo, SP, Brazil}
\author{Marcelo G.~Munhoz\\Instituto de F\'{i}sica, Universidade de S\~{a}o Paulo, C.P. 66318, 05315-970 S\~{a}o Paulo, SP, Brazil}

\abstract{
Recent experimental data at the Large Hadron Collider (LHC) confirmed that the soft sector azimuthal anisotropies in central XeXe collisions are sensitive to deformations in the wave function of the Xenon nucleus. Additionally, the CMS experiment found that D meson flow is slightly suppressed compared to other particle species when considering quark number scaling in small systems of pPb compared to PbPb. In this talk we used the D and B Mesons Modular (DAB-MOD) code coupled to Trento+v-USPhydro to calculate the $R_{AA}$ and $v_n$’s of D mesons using different energy loss models and Langevin techniques. Comparing D mesons $R_{AA}$ and $v_n$ in PbPb to XeXe collisions it is found that in central collisions D meson azimuthal anisotropies are sensitive to details in the nuclear structure wave function of Xenon.  Additionally, we find that in mid-central collisions the smaller system size of XeXe suppresses D meson flow.
}

\FullConference{13th International Workshop in High pT Physics in the RHIC and LHC Era (High-pT2019)\\
		19-22 March 2019\\
		Knoxville, Tennessee, USA}

\begin{document}

\section{Introduction}

Two of the crucial signals of the Quark Gluon Plasma (QGP) - the most perfect fluid known to humanity - are collective flow and energy loss.  Collective flow of long range correlations has been measured via the Fourier coefficients of the particle spectra i.e. $v_n\{m\}$ where n indicates the harmonic and m the number of particles correlated. While a finite  $v_2\{m\}$ arises primarily from the geometry of the overlap between two heavy ions, the measured finite $v_3\{m\}$ exists because of  the quantum mechanical fluctuations of the position of the nucleons in the overlap region \cite{Alver:2010gr,Takahashi:2009na}. Energy loss is measured via the nuclear modification factor $R_{AA}$ where the ratio of the particle spectra in a large AA system is normalized by the particle spectra in a pp collision all normalized by the number of binary collisions.  When no energy loss is experienced $R_{AA}\rightarrow 1$ otherwise for a suppression   $R_{AA}< 1$ (for a review see \cite{Jacobs:2004qv}).

The dynamics of the soft sector of the strongly interacting soup of quarks and gluons can be described very well via event-by-event relativistic viscous hydrodynamical models that assume an almost vanishing shear viscosity to entropy density ratio \cite{Niemi:2015voa,Noronha-Hostler:2015uye,Eskola:2017bup,Giacalone:2017dud,Gale:2012rq,Bernhard:2016tnd}.  For the hard and heavy flavor sector additional energy loss mechanisms are needed in order to describe the suppression of both high transverse momentum, $p_T\gtrapprox 10$ GeV and heavy quarks particles as they lose energy traveling through the dense QGP. Additionally, it was found in recent years that realistic hydrodynamical medium was required in order to describe the azimuthal anistropies of these hard probes \cite{Noronha-Hostler:2016eow,Betz:2016ayq} (especially $v_3$ and multiparticle cumulants \cite{Sirunyan:2017pan}).

When the QGP was first discovered, heavy AA collisions were explored because it was thought that only a large system would be able to produce the QGP.  However, in recent years it has been found by a variety of experiments \cite{Aaboud:2017acw,Aaboud:2017blb,Aad:2013fja,Sirunyan:2018toe,Chatrchyan:2013nka,Khachatryan:2014jra,Khachatryan:2015waa,Khachatryan:2015oea,Sirunyan:2017uyl,ABELEV:2013wsa,Abelev:2014mda,Adare:2013piz, Adare:2014keg,Aidala:2018mcw,	Adare:2018toe,	Adare:2015ctn,	Aidala:2016vgl,	Adare:2017wlc,	Adare:2017rdq,	Aidala:2017pup,	Aidala:2017ajz} signals for the QGP in small systems such as pPb, dAu, $^3$HeAu, pAu, and possibly even pp, which have been reasonably well reproduced and predicted by relativistic hydrodynamics \cite{Bozek:2011if,Bozek:2012gr,Bozek:2013ska,Bozek:2013uha,Kozlov:2014fqa,Zhou:2015iba,Zhao:2017rgg,Mantysaari:2017cni,Weller:2017tsr} (although a number of questions still remain). At the same time in small systems the nuclear modification factor was equivalent with 1 \cite{Khachatryan:2016odn}, which implied that no energy loss was seen or that the definitions of energy loss had to be refined in small systems. However, a surprising recent result was then released by the CMS collaboration in pPb collisions where they found that the heavy flavor sector still produced a significant $v_2$ but that a suppression was seen compared to other lighter particles \cite{Sirunyan:2018toe}. 

Because of these recent results from CMS, we use D mesons as a test bed for energy loss in small systems.  In this proceedings we explore the sensitivity of D mesons to the system size, specifically making the comparison between PbPb and XeXe collisions, which were recently ran at the LHC.  However, XeXe collisions provided a surprise beyond system size effects.  Assuming a spherical Xenon nucleus, hydrodynamic predictions of $v_2\{2\}$ in central collision for the ratio of XeXe/PbPb collision significantly under-predicted the experimental data \cite{Giacalone:2017dud,Eskola:2017bup}, however, the inclusion of a deformed Xenon nucleus shifted the theoretical predictions closer to the experimental data \cite{Giacalone:2017dud}. Since this is an effect that is only seen in central collisions, a PbPb to XeXe comparison allows a test of both system size dependence (in the $30-50\%$ centrality class) and the sensitivity to a deformed nucleus in the heavy flavor sector (in the $0-10\%$ centrality class).  

\section{Model}

In recent years it was found that a more realistic description of the QGP medium is a necessary tool for simultaneously  reproducing the nuclear modification factor $R_{AA}$ and the high $p_T$ and heavy flavor azimuthal anisotropies $v_n\{m\}(p_T)$ \cite{Nahrgang:2014vza,Noronha-Hostler:2016eow,Betz:2016ayq,Prado:2016szr}.  Therefore we include a realistic event-by-event description of the medium coupled to modular heavy flavor code using Trento+v-USPhydro+DAB-MOD.  The initial conditions are generated using  Trento \cite{Moreland:2014oya} where $p=0$, $\sigma=0.51$ fm, and $k=1.6$ for PbPb 5.02 TeV and for XeXe 5.44 TeV (note for Xenon we use the deformed Wood-Saxon parameters described in \cite{Moller:2015fba,Giacalone:2017dud}). The hydrodynamic model is v-USPhydro \cite{Noronha-Hostler:2013gga,Noronha-Hostler:2014dqa,Noronha-Hostler:2015coa} where we use the same parameters for LHC run 2 as in \cite{Alba:2017hhe} where $\tau_0=0.6$ fm, $\eta/s\sim 0.05$, and $T_{FO}=150$ MeV. To describe the heavy flavor sector we run DAB-MOD \cite{Prado:2016szr} a modular Monte Carlo simulation that samples heavy quarks  using distributions from pQCD FONLL calculations~\cite{Cacciari:1998it,Cacciari:2001td}, implements either a parameterized energy loss model+energy loss fluctuations or a Langevin model, and then this is followed by fragmentation functions and coalescence to obtain the final particle yields. 

In order to make predictions for smaller systems size we hold all our parameters fixed based on the PbPb 5.02 TeV results, which have already been well-tested compared to experimental data \cite{Katz:2018vvm}. Then the system size dependence arises through the initial conditions+hydrodynamical backgrounds calculated within Trento+v-USPhydro that were already discussed in detail in \cite{Sievert:2019zjr}.

\section{Deformed Xe}

The geometry of the initial conditions can be quantified using the vector quantities of eccentricities 
\begin{equation}
\mathcal{E}_n=\frac{\int r^n e^{in\phi}\rho(r,\phi) r drd\phi}{\int r^n \rho(r,\phi) r drd\phi},
\end{equation}
which are very strongly correlated with the final flow harmonics vector on an event-by-event basis in the soft sector \cite{Teaney:2010vd,Gardim:2011xv,Niemi:2012aj,Teaney:2012ke,Qiu:2011iv,Gardim:2014tya}, high $p_T$ all charged particles \cite{Betz:2016ayq,Noronha-Hostler:2016eow}, and the heavy flavor sector \cite{Nahrgang:2014vza,Prado:2016szr}. We will mostly be focusing on only the magnitude of the eccentricities, $\varepsilon_n$, in this paper due to the changes it experiences for deformed nucleus as well as system size dependence. 

\begin{figure}[!htb] 
  \centering
 \includegraphics[width=\textwidth]{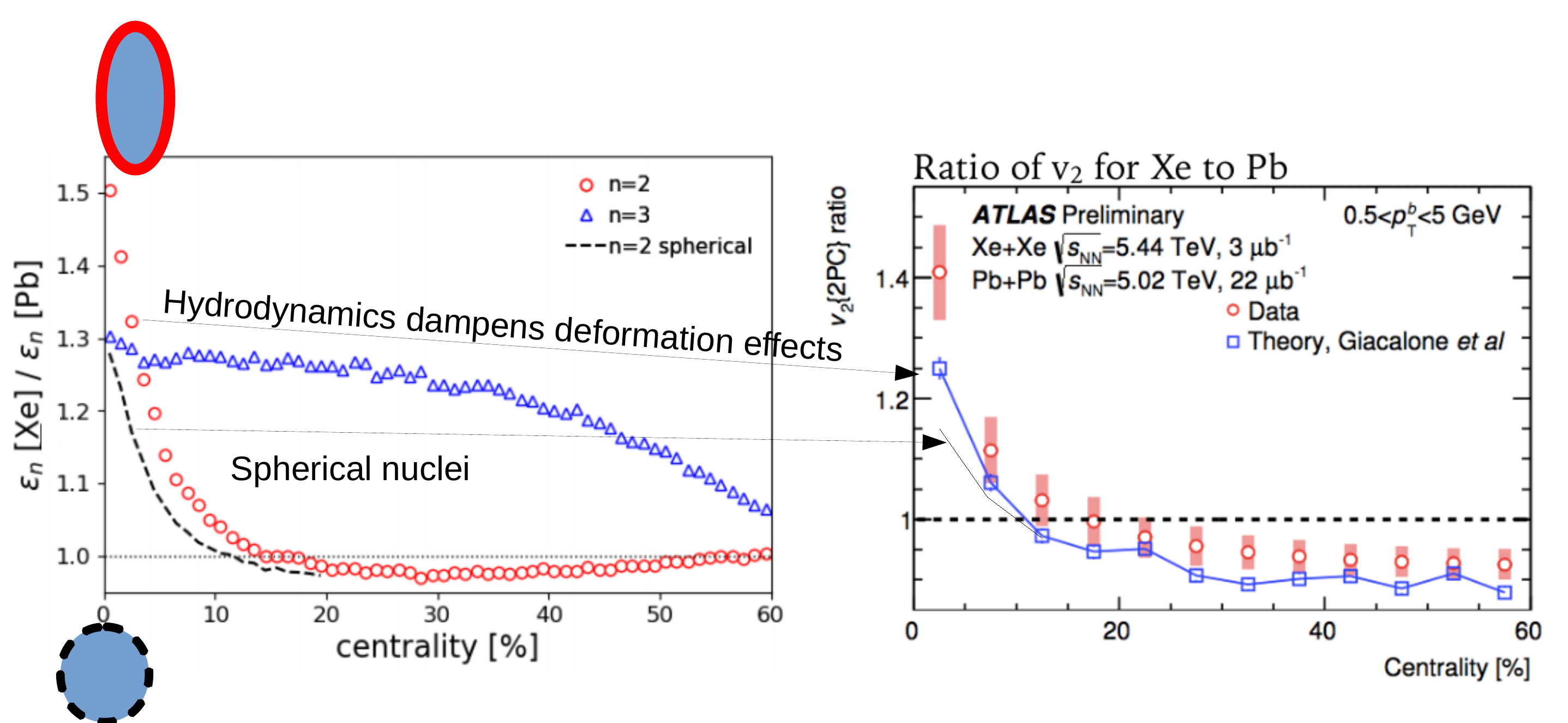}
  \caption{  On the left, ratio of the eccentricities of Pb Pb/Xe Xe collisions at the LHC calculated using TRENTO \cite{Moreland:2014oya}. Figure taken from \cite{Giacalone:2017dud}.  On the right, theory predictions from trento+v-USPhydro with and without a deformed Xe nucleus compare to experimental data from \cite{ATLAS:2018iom}  }
  \label{fig:softXE}
\end{figure}

In Fig.\ \ref{fig:softXE} we compare the effect of the ratio of the eccentricities in XeXe to PbPb for $\varepsilon_2$ and $\varepsilon_3$. For spherical AA collisions we generally expect that in central collisions $\varepsilon_2$ and $\varepsilon_3$ are inversely proportional to the system size (see Fig. 10 from \cite{Sievert:2019zjr}).  Since the radius of Xenon is smaller than Lead we generally expect that the $v_n$'s will also be larger, which is seen in the right plot in  Fig.\ \ref{fig:softXE} for both the experiment and theory.  However, what is also clear is that the prediction for a spherical Xenon nucleus significantly underpredicts the experimental data for $v_2^{Xe}\{2\}/v_2^{Pb}\{2\}$ in central collisions.  The theoretical prediction for a deformed Xenon nucleus is closer to experimental data (CMS \cite{CMS:2018jmx}, ALICE \cite{Acharya:2018ihu}, and ATLAS \cite{ATLAS:2018iom}) but still slightly somewhat undepredicts the effect \cite{Giacalone:2018cuy}, it may be that a larger deformation is required or that other nuclear structure effects are missing from heavy-ion initial condition models. Finally, it should be clear that the effect of the deformation is only relevant in central collisions.  For more peripheral collisions the dominating geometrical effect comes from the impact region itself. 

\begin{figure}[!htb] 
  \centering
 \includegraphics[width=\textwidth]{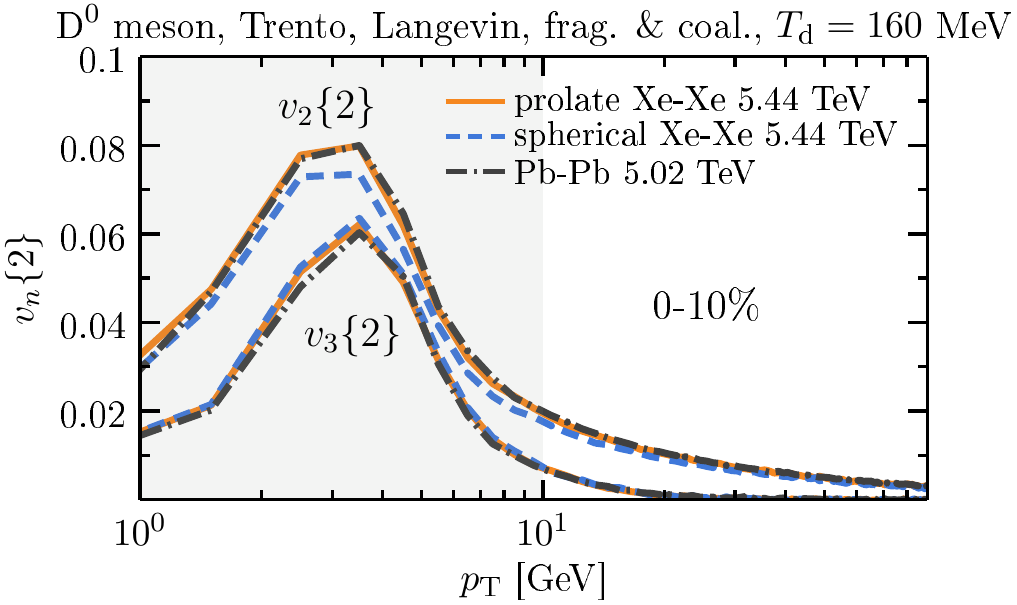}
  \caption{  $v_n\{2\}(p_T)$ of D mesons in $0-10\%$ PbPb vs. XeXe collisions compared using Trento+v-USPhydro+DAB-MOD. }
  \label{fig:defv2v3}
\end{figure}

In Fig.\ \ref{fig:defv2v3} we compare $v_2\{2\}(p_T)$ and $v_3\{2\}(p_T)$ from PbPb 5.02 TeV collisions to both a spherical Xenon nucleons versus a prolate Xenon for XeXe 5.44 TeV collisions in the $0-10\%$ centrality class.  We find a small enhancement in $v_3\{2\}$, which is consistent with the soft sector results from \cite{Giacalone:2017dud,Sievert:2019zjr}.  This enhancement arises due to the universal scaling of $v_3\{2\}$ with Npart as demonstrated in  Fig.\ 17 from \cite{Sievert:2019zjr}. Since $v_3\{2\}$ has a maximum in $N_{part}\approx 100$ and central XeXe collisions have an $N_{part}\approx200-250$ compared to PbPb's $N_{part}\approx310-415$ then one expects a continually growing $v_3\{2\}$ as $N_{part}\rightarrow 100$. We note that the $v_3\{2\}$  result has nothing to do with a deformed nucleus since both a prolate and spherical nucleus are in agreement.  

More interestingly, however, is $v_2\{2\}$ that shows a clear enhancement for a prolate Xenon nucleus whereas for a spherical Xenon nucleus one predicts that it should be suppressed at intermediate $p_T$'s.  This is a quite interesting result because it is not even happening in just the ultracentral collisions where the original XeXe effect of deformation was most sensitive  \cite{Giacalone:2017dud} but rather in the $0-10\%$ centrality class.  Additionally, it appears that intermediate $p_T$ ($p_T=2-5$ GeV) is the most sensitive to the deformation, which is also surprising.  

\section{System size: PbPb vs. XeXe collisions}

\begin{figure}[!htb] 
  \centering
 \includegraphics[width=\textwidth]{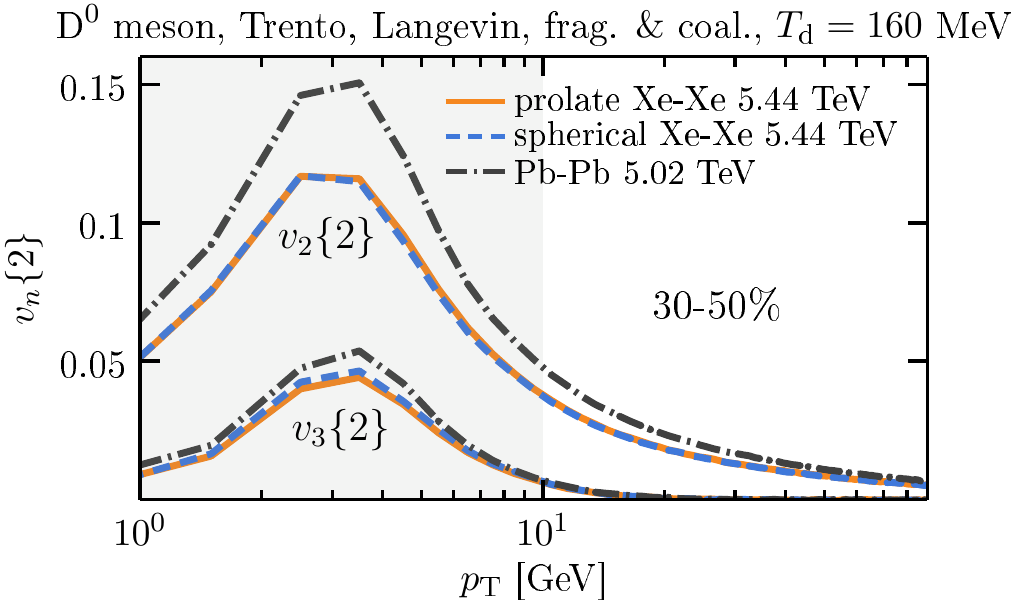}
  \caption{  $v_n\{2\}(p_T)$ of D mesons in $30-50\%$ Pb Pb vs. Xe Xe collisions compared using Trento+v-USPhydro+DAB-MOD. }
  \label{fig:3050}
\end{figure}

Returning to our original purpose of studying the system size dependence of D mesons $v_n\{2\}(p_T)$ we now compare PbPb collisions to XeXe collisions.  In Fig.\ \ref{fig:softXE} it was shown that the eccentricties are identical for the spherical and deformed Xenon nucleus for any collisions in centralities more peripheral than $\sim 20\%$.  Thus, exploring comparisons in the $30-50\%$ centrality class is certainly safe from deformation effects. Indeed, in Fig.\ \ref{fig:3050} we compare both $v_2\{2\}(p_T)$ and $v_3\{2\}(p_T)$ from PbPb collisions and XeXe collisions and see that there is no significant difference between the prolate and spherical Xenon nucleus.  

From the eccentricities shown in Fig.\ \ref{fig:softXE} we expect that $v_2^{Pb}\{2\}(p_T)>v_2^{Xe}\{2\}(p_T)$ and that $v_3^{Pb}\{2\}(p_T)\lesssim v_3^{Xe}\{2\}(p_T)$, which is what is found in the soft sector \cite{Giacalone:2018cuy}.  While the prediction from the eccentricities for $v_2$ holds for D mesons, we find that $v_3^{Pb}\{2\}(p_T)> v_3^{Xe}\{2\}(p_T)$, which is in contrast to both the eccentricities predictions and what is seen in the soft sector.  There are a number of reasons why this may occur, for instance, the higher decoupling temperature that we use in the heavy flavor sector or the possibility that there is a nontrivial interplay between $v_3$ with the physics of the heavy flavor sector i.e.\ energy loss, fragmentation functions, or coalescence. However, we also note that this suppression of the expected $v_n$'s in smaller systems is consistent with what CMS found comparing pPb to PbPb collisions.  Thus, we are encouraged that the heavy flavor sector can provide new information on the question of energy loss in small systems. 

\section{Conclusions}

In conclusion, we have made the first event-by-event azimuthal anisotropies predictions  in XeXe 5.44 TeV collisions (note previous predictions were made for XeXe collision on top of a non-fluctuating background using a Bjoerken expansion in \cite{Zigic:2018ovr,Djordjevic:2018ita}). We find that surprisingly enough that D mesons are sensitive to a deformed $^{129}Xe$ in the $0-10\%$ centrality class  and that this effect is enhanced in the range of $p_T\sim 3-5$ GeV. From this we conclude that it would be very interesting for experimentalists to compare D meson azimuthal anisotropies in deformed AA collisons (such as Uranium or the upcoming isobar run) as well. 

In mid-central collisions of $30-50\%$ centrality we have shown that there is a suppression of the $v_n\{2\}(p_T)$'s due to system size effects.  From the eccentricties alone, one would expect a small suppression  on the order of a few percentage points such that $v_2^{Xe}\{2\}<v_2^{Pb}\{2\}$ in mid-central collisions but that cannot explain the over $20\%$ suppression of D mesons $v_2^{Xe}\{2\}$ compared to $v_2^{Pb}\{2\}$.  Thus it appears that D mesons are more sensitive to system size effects. 

We have not yet checked the sensitivity to other medium parameters such as $\tau_0$ \cite{Andres:2019eus} and $\eta/s$ in peripheral collisions \cite{Betz:2016ayq}. It also remains to be seen if the hard and heavy flavor sector are sensitive to the equation of state as explored in \cite{Alba:2017hhe}, which could be connected to the initialization time as well. 

\section*{Acknowledgments} 
 
The authors thank Funda\c{c}\~ao de Amparo \`a Pesquisa do Estado de S\~ao Paulo (\textsc{fapesp}) and Conselho Nacional de Desenvolvimento Cient\'ifico e Tecnol\'ogico (\textsc{cnp}q) for support. C.A.G.P. is supported by the NSFC under grant No. 11521064, MOST of China under Project No. 2014CB845404.
J.N.H. acknowledges the support of the Alfred P. Sloan Foundation, support from the US-DOE Nuclear Science Grant
No. DE-SC0019175, and the Office of Advanced Research Computing (OARC) at Rutgers, The
State University of New Jersey for providing access to the Amarel cluster and associated research
computing resources that have contributed to the results reported here. JN is partially supported by CNPq grant 306795/2017-5 and FAPESP grant 2017/05685-2.

\bibliographystyle{JHEP}
\bibliography{BIG}

\end{document}